\documentclass[floatfix,showpacs,showkeys,aps,prl,twocolumn]{revtex4-2}
\usepackage{lineno}
\usepackage{amsmath}
\usepackage{amssymb}

\usepackage{float}
\usepackage{tikz}
\usepackage{lipsum,adjustbox}
\usepackage{graphicx}
\usepackage[caption = false]{subfig}
\usepackage{dcolumn, multirow, booktabs, siunitx, rotating, array}
\usepackage{bm}
\usepackage{gensymb}

\usepackage{tabstackengine}
\usepackage[inline]{enumitem}

\setstackEOL{\cr}
\setstackgap{L}{\normalbaselineskip}

\let\oldequation\equation
\let\oldendequation\endequation

\renewenvironment{equation}
  {\linenomathNonumbers\oldequation}
  {\oldendequation\endlinenomath}

\newcolumntype{L}[1]{>{\raggedright\arraybackslash}p{#1}}
\newcolumntype{C}[1]{>{\centering\arraybackslash}p{#1}}
\newcolumntype{R}[1]{>{\raggedleft\arraybackslash}p{#1}}

\usepackage{hyperref}
\hypersetup{
 pdftitle={},
 pdfauthor={},
 pdfsubject={},
 pdfkeywords={},
 pdfstartview={},
 bookmarksopen=true, breaklinks=true, debug=true,
 colorlinks=true, linkcolor=blue, citecolor=red, urlcolor=blue,
 hyperfigures=true
}

\usepackage{todonotes}
\newcommand{\BESIIIorcid}[1]{\href{https://orcid.org/#1}{\hspace*{0.1em}\raisebox{-0.45ex}{\includegraphics[width=1em]{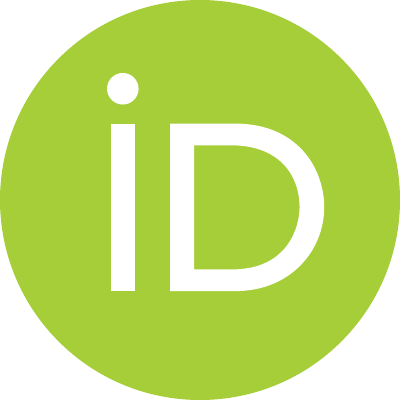}}}}

\begin{document}

\title{\boldmath Unexpected Large Relative Strong Phase and Search for Isospin Breaking and $CP$ Asymmetries in $J/\psi \to K^*(892)\bar K$}
\author{
M.~Ablikim$^{1}$\BESIIIorcid{0000-0002-3935-619X},
M.~N.~Achasov$^{4,c}$\BESIIIorcid{0000-0002-9400-8622},
P.~Adlarson$^{83}$\BESIIIorcid{0000-0001-6280-3851},
X.~C.~Ai$^{89}$\BESIIIorcid{0000-0003-3856-2415},
C.~S.~Akondi$^{31A,31B}$\BESIIIorcid{0000-0001-6303-5217},
R.~Aliberti$^{39}$\BESIIIorcid{0000-0003-3500-4012},
A.~Amoroso$^{82A,82C}$\BESIIIorcid{0000-0002-3095-8610},
Q.~An$^{78,65,\dagger}$,
Y.~H.~An$^{89}$\BESIIIorcid{0009-0008-3419-0849},
M.~S.~Anderson$^{39}$\BESIIIorcid{0009-0008-1550-2632}
% ...
\\
\vspace{0.2cm}
(BESIII Collaboration)\\
\vspace{0.2cm} {\it
$^{1}$ Institute of High Energy Physics, Beijing 100049, People's Republic of China\\
% ...
$^{4}$ Budker Institute of Nuclear Physics SB RAS (BINP), Novosibirsk 630090, Russia\\
% ...
$^{31}$ INFN Sezione di Ferrara, (A)INFN Sezione di Ferrara, I-44122, Ferrara, Italy; (B)University of Ferrara, I-44122, Ferrara, Italy\\
% ...
$^{39}$ Johannes Gutenberg University of Mainz, Johann-Joachim-Becher-Weg 45, D-55099 Mainz, Germany\\
% ...
$^{65}$ State Key Laboratory of Particle Detection and Electronics, Beijing 100049, Hefei 230026, People's Republic of China\\
% ...
$^{78}$ University of Science and Technology of China, Hefei 230026, People's Republic of China\\
% ...
$^{82}$ University of Turin and INFN, (A)University of Turin, I-10125, Turin, Italy; (B)University of Eastern Piedmont, I-15121, Alessandria, Italy; (C)INFN, I-10125, Turin, Italy\\
$^{83}$ Uppsala University, Box 516, SE-75120 Uppsala, Sweden\\
% ...
$^{89}$ Zhengzhou University, Zhengzhou 450001, People's Republic of China\\
\vspace{0.2cm}
$^{\dagger}$ Deceased\\
% ...
$^{c}$ Also at the Novosibirsk State University, Novosibirsk, 630090, Russia\\
% ...
}}

\date{\today}

\begin{abstract}
Using a direct scan of 26 energy points near the $J/\psi$ resonance, we perform the first measurement of the relative phase $\phi_{\gamma, 3g}$ between the strong and electromagnetic amplitudes in $J/\psi\to\bar K^0 K^*(892)^0+\mathrm{c.c.}$ and $J/\psi\to K^+ K^*(892)^-+\mathrm{c.c.}$. Unexpectedly, the phase in the neutral channel is found to be $\sim 150^\circ$, deviating from orthogonality ($90^\circ$) by $4.8\sigma$ and from a relative real amplitude ($0^\circ$ or $180^\circ$) by $12.6\sigma$ or $2.8\sigma$, respectively. By contrast, the charged channel phase is consistent with $\sim 180^\circ$ within $1\sigma$, exhibiting channel-dependent behavior. The corresponding branching fractions are consistent with the world averages but achieve better than twofold improvement in precision. The ratios between the branching fractions of $J/\psi\to\bar K^0 K^*(892)^0+\mathrm{c.c.}$ and $J/\psi\to K^+ K^*(892)^-+\mathrm{c.c.}$ are also measured. After subtracting the electromagnetic contribution, the corresponding strong contribution ratios obey isospin symmetry within $1.4\sigma$. A search for direct $CP$ violation yields asymmetries consistent with zero.
\end{abstract}

\maketitle

Quantum Chromodynamics (QCD) is well tested in the perturbative region in high energy physics~\cite{QCD1,QCD2,QCD3,QCD4}. However, the dynamics in the transition region between the perturbative and nonperturbative regions where a heavy $c\bar c$ pair hadronizes into light-hadron final states remains poorly understood~\cite{QCDheavyquark,Quarkonia,WangXFrequire}. Systematic studies of charmonium decays are therefore essential for clarifying how perturbative QCD (pQCD) connects to nonperturbative dynamics and for improving our understanding of light-hadron hadronization. More than 50 years after the discovery of the $J/\psi$, its decay dynamics remain incompletely resolved. This is reflected in its unusually narrow width and in the fact that its dominant decay mechanisms are the Okubo-Zweig-Iizuka (OZI)-suppressed strong three-gluon amplitude and the electromagnetic (EM) one-photon amplitude. Long-standing anomalies, including the ``$\rho\pi$ puzzle'' (Mark II found $R_{\rho\pi}<0.6\%$), the apparent violation of the pQCD ``12\% rule''~\cite{rhopiMarkII}, and the observed $\sim 90^\circ$ relative phase between the three-gluon and one-photon amplitudes~\cite{ppphase,NNphase}, indicate sizable absorptive effects beyond a purely short-distance description~\cite{nuFF,heliSele}. Many theoretical models have been proposed to address these tensions.

The phenomenological SU(3) flavor symmetry model can reproduce such $90^\circ$ phase, while the dynamical origin of a large absorptive contribution is still unknown~\cite{SU3-1-inter,SU3-2-jpsietac,SU3-3-mo}. Possible sources of an imaginary contribution include mixing between the $J/\psi$ and a vector glueball~\cite{gluonball1,gluonball2,gluonball3} or hadronic loops and final-state interactions (FSI)~\cite{FSI1,FSI2}. The glueball mixing scenario naturally yields a nearly channel-independent imaginary amplitude and provides a potential explanation of the ``$\rho\pi$ puzzle.'' Alternatively, it has been argued that OZI-violating mechanisms alone can generate a predominantly imaginary amplitude once the running of $\alpha_s$ is treated in a dispersive framework~\cite{OZIamp}, without invoking an additional gluonium component. In another approach, the $J/\psi$ couples to off-shell $D^{(*)}\bar D^{(*)}$ intermediate states~\cite{FSIopechar1,FSIopechar2}, which naturally carry strong phases, leading to channel-dependent interference. Moreover, the mass splittings between charged and neutral charmed mesons ($m_{D^{(*)+}}\neq m_{D^{(*)0}}$) can enhance isospin-violating transitions, which has been used to explain the enhancement of $\psi(2S)\to\pi^0 J/\psi$ relative to $\psi(2S)\to\eta J/\psi$~\cite{pi0jpsi}. Therefore, isospin symmetry may also be broken in $J/\psi$ decays. These strong phases also motivate direct $CP$ tests. Within the standard model, purely strong and QED charmonium decays are expected to exhibit negligible direct $CP$ violation, with weak interactions entering only at the $\mathcal{O}(10^{-8})$ level~\cite{Jpsiweak}. Thus any percent-level $CP$ asymmetries would be anomalous, and precision comparisons of charge-conjugate modes provide clean null tests. Therefore, a direct determination of $\phi_{\gamma,3g}$ from the energy-dependent interference together with precision tests of isospin symmetry and $CP$ invariance in exclusive $J/\psi \to K^*(892)\bar K$ decays is urgently needed to discriminate among these scenarios.

In this Letter, we study the cross-section line shape of $e^+e^-\to K_S^0K^+\pi^-$ using BESIII data collected at 26 center-of-mass (c.m.) energies between 3000.00 and $3119.88~\text{MeV}$, corresponding to a total integrated luminosity of $440.7~\text{pb}^{-1}$~\cite{Abakumova:2011rp,jlumi,jpsinumber}. With a partial wave analysis (PWA), we isolate the contributions from the dominant subprocesses $e^+e^-\to \bar K^0K^*(892)^0$ and $e^+e^-\to K^+K^*(892)^-$. Throughout this Letter, charged conjugations are always implied except when discussing $CP$ tests.

The total Born cross section for $e^+e^-\to f$ is expressed as a coherent sum of resonance and continuum amplitudes~\cite{jpsidecay}:
\begin{equation}
  \sigma_\text{Born}^{f}(s) \propto \left|\mathcal{A}^{f}_\text{cont}(s)+\mathcal{A}^{f}_\gamma(s)+\mathcal{A}^{f}_{3g}(s) \right|^2.
\end{equation}
Assuming that the three contributions share the same effective final-state dynamics, the Born cross section is written as~\cite{wangpingformular,wangyadiformular,Zlineshape,kedr}
\begin{widetext}
  \begin{equation}\label{eq:xsformular}
    \sigma_\text{Born}^{f}=\frac{4 \pi \alpha^2}{3 s}\left|1+(1+\mathcal{C} e^{i \phi_{\gamma, 3g}}) \frac{s}{M} \frac{3 \sqrt{\Gamma^0_{e e} \Gamma^0_{\mu\mu}} / \alpha}{s-M^2+i M \Gamma}e^{i\phi_{\gamma,\text{cont}}}\right|^2  \mathcal{P}(s)  \frac{\mathcal{F}^2}{s^{n}},
  \end{equation}
\end{widetext}
where $s$ is the square of the c.m. energy; $\alpha$ is the fine-structure constant; $\Gamma_{ee}^0$ and $\Gamma_{\mu\mu}^0$ are the bare electronic and muonic widths, respectively; $M$ and $\Gamma$ are the mass and width of the $J/\psi$ resonance, respectively; $\mathcal{C}$ is the ratio $\frac{\left|\mathcal{A}_{3g}\right|}{\left|\mathcal{A}_\gamma\right|}$; and $\phi_{\gamma,3g(\rm cont)}$ is the relative phase between strong (continuum) and EM amplitudes. The factor $\mathcal{F}^2 / s^n$ parameterizes the squared effective timelike EM form factor for the exclusive final state $f$, with $n$ governing the energy dependence. The symbol $\mathcal{P}(s)$ denotes the dimensionless phase-space (PHSP) factor. Given the narrow width of the $J/\psi$ resonance, the observed cross section is modeled by applying the vacuum polarization correction and convolving with both initial-state radiation (ISR) function and the beam energy spread distribution,
\begin{equation}\label{eq:fitformular}
  \begin{aligned}
  \sigma_{\rm obs}(\sqrt{s})= &
  \int_{\sqrt{s}-5S_E}^{\sqrt{s}+5S_E} \frac{d\sqrt{s'}}{\sqrt{2\pi}S_E} e^{-\frac{(\sqrt{s}-\sqrt{s'})^2}{2S_E^2}}\\
  &\int_{0}^{1-\frac{s_{\min}}{s'}}dx F_{\rm ISR}(x,s') \frac{\sigma_\text{Born}^{f}(s'(1-x))}{\left|1-\Pi_0(s'(1-x))\right|^2},\\
  \end{aligned}
\end{equation}
where $F_{\rm ISR}$ is the Kuraev-Fadin radiator~\cite{ISRfactor}, $\Pi_0$ is the nonresonant vacuum polarization operator $\Pi$~\cite{VPfactor}, $x$ is the fraction of the c.m. energy lost due to ISR photon, $S_E$ is the beam energy spread, and $s_{\min}=(2.85~\text{GeV})^2$.

Based on a $\chi^2$ line shape fit~\cite{chifit} to Eq.~\eqref{eq:fitformular}, we determine the relative magnitude and phase of the strong and EM amplitudes and the branching fraction (BF) in $J/\psi \to K^0_SK^+\pi^-$. In combination with the PWA results, these parameters are extracted in its dominant subprocesses $J/\psi\to K^+ K^*(892)^-$ and $J/\psi\to\bar K^0 K^*(892)^0$. Isospin breaking is quantified by the neutral-to-charged ratios at the Born level, $\mathcal{R}_\sigma = \sigma_{\rm Born}(e^+e^-\to\bar K^0 K^*(892)^0)/\sigma_{\rm Born}(e^+e^-\to K^+ K^*(892)^-)$ and in the BF ratio, $\mathcal{R}_{K^*\bar{K}}=\mathcal{B}(J/\psi\to\bar K^0K^*(892)^0)/\mathcal{B}(J/\psi\to K^+K^*(892)^-)$. To separate strong and EM contributions, additional fraction factors are defined as $f_{3g} = |\mathcal{A}_{3g}|^2/|\mathcal{A}_{3g} + \mathcal{A}_\gamma|^2 = \mathcal{C}^2/(1 + 2\mathcal{C}\cos\phi_{\gamma,3g} + \mathcal{C}^2)$ and $f_{\gamma} = |\mathcal{A}_\gamma|^2/|\mathcal{A}_{3g} + \mathcal{A}_\gamma|^2=1/(1+2\mathcal{C}\cos\phi_{\gamma,3g}+\mathcal{C}^2)$.
For each channel $(i)$, we obtain $f_{3g}^{(i)}$ and $f_{\gamma}^{(i)}$ from the best-fit parameters. The neutral-to-charged ratios for the pure strong and pure EM parts are then $\mathcal{R}^{3g}_{K^*\bar{K}} = \mathcal{R}_{K^*\bar{K}}\times f_{3g}^{(K^*(892)^0)}/f_{3g}^{(K^*(892)^-)}$ and $\mathcal{R}^{\gamma}_{K^*\bar{K}} = \mathcal{R}_{K^*\bar{K}}\times f_{\gamma}^{(K^*(892)^0)}/f_{\gamma}^{(K^*(892)^-)}$, respectively.
Direct $CP$ null tests are performed using the Born cross sections, $A_{\sigma_{\rm Born}}=(\sigma^+_{\rm Born}-\sigma^-_{\rm Born})/(\sigma^+_{\rm Born}+\sigma^-_{\rm Born})$, where the superscripts ``$+$'' and ``$-$'' denote the samples split by the unambiguous flavor tag provided by the charged kaon. For each channel we determine asymmetries of the fit parameters: $A_\mathcal{C}=(\mathcal{C}^+ - \mathcal{C}^-)/(\mathcal{C}^+ + \mathcal{C}^-)$, $A_\mathcal{B}=(\mathcal{B}^+ - \mathcal{B}^-)/(\mathcal{B}^+ + \mathcal{B}^-)$, and $\Delta\phi_{\gamma,3g}=\phi_{\gamma,3g}^{+}-\phi_{\gamma,3g}^{-}$. These observables probe possible $CP$ asymmetries in the relative magnitude, the BF, and the phase, respectively.

A description of the design and the performance of the BESIII detector can be found in Refs.~\cite{BES3,Yu:2016cof}. The inclusive Monte Carlo (MC) sample, described in Refs.~\cite{geant4,kkmc,evtgen,lundcharm,photos}, has been validated for background simulation. The signal decays $J/\psi\to K_S^0 K^+\pi^-$, $J/\psi\to\bar K^0 K^*(892)^0$, and $J/\psi\to K^+ K^*(892)^-$ are generated with ConExc~\cite{conexc} using the helicity amplitude with parameters fixed to the PWA results obtained in this Letter.

The selection criteria of $\pi^\pm$, $K^\pm$, and $K_S^0$ candidates are detailed in Ref.~\cite{prdversion}. A four-constraint (4C) kinematic fit imposing energy-momentum conservation is performed, and candidate events are required to satisfy $\chi_\text{4C}^2<200$. To remove gamma conversion Bhabha background, the opening angle between the two pions from $K_S^0$ decay is required to be greater than 30$^\circ$. The remaining background contribution is evaluated using the $J/\psi$ inclusive MC and exclusive MC samples of $e^+e^-\to K_S^0K^+\pi^-\pi^0$, $e^+e^-\to K^+K^-\pi^+\pi^-$, and $e^+e^-\to 2(\pi^+\pi^-)$. After considering the respective cross sections~\cite{BaBarkkpipi, BaBarkskpipi0, BaBar4pi, pdg}, the background level is estimated to be 0.2\%, indicating a negligible background contribution. Consequently, the signal yield for $e^+e^-\to K_S^0 K^+ \pi^-$ is obtained simply by counting events. To improve the mass resolution, a five-constraint kinematic fit is performed by imposing an additional constraint that the invariant mass $M_{\pi^+\pi^-}$ be equal to the nominal $K_S^0$ mass~\cite{pdg}.

\begin{figure*}[htbp]\centering
  \includegraphics[width=0.99\textwidth]{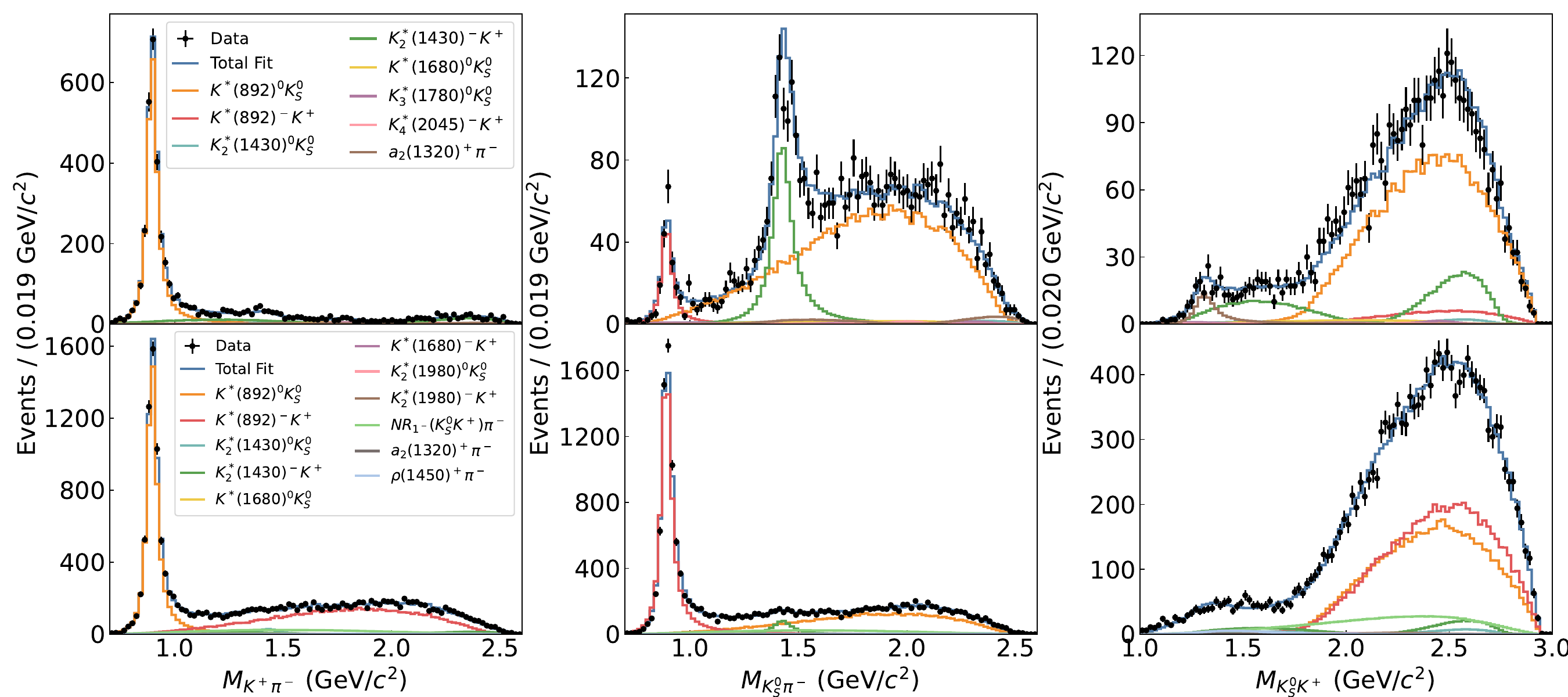}
  \caption{Projections of the PWA fits on mass spectra for the dataset taken at $\sqrt s=3080.00$ MeV (top) and at $\sqrt s=3096.20$ MeV (bottom).}
	\label{fig:pwaresult}
\end{figure*}

A PWA fit is performed on the surviving candidate events to identify the intermediate processes. The decay amplitude is constructed using the helicity amplitude formalism~\cite{Chungformalism, Richmanformalism}, and the full fit procedure is implemented with the TF-PWA framework~\cite{tfpwajiangyi}. The fundamental concepts are detailed in the Physical Review D companion paper~\cite{prdversion}.

The amplitude of a complete chain is constructed as the product of each two-body amplitude and the resonant propagator $R$. The propagators of all resonances are described by the relativistic Breit-Wigner formula~\cite{rBW, centrifugal}. For the nonresonant (NR) contribution, the dynamical part is set to unity. All possible intermediate states from the Particle Data Group (PDG)~\cite{pdg} that satisfy $J^{PC}$ conservation are considered. The PWA fit procedure starts by including $K^*(892)^0$, $K^*(892)^-$, $K_2^*(1430)^0$, $K_2^*(1430)^-$, and $a_2(1320)^+$ as the initial baseline solution. For the resonances, masses and widths are fixed to PDG values~\cite{pdg}.

To determine the set of contributing intermediate states, resonance scans are performed at two reference energy points, $\sqrt{s}=3080.00$ and $3096.20~\text{MeV}$, chosen for their high integrated luminosities and for representing distinct $e^+e^-$ annihilation regimes: the continuum-dominated process $\mathcal{A}_\text{cont}$ at $3080.00~\text{MeV}$, and the resonance region where $\mathcal{A}_\text{cont}, \mathcal{A}_\gamma, \mathcal{A}_{3g}$ interfere at $3096.20~\text{MeV}$. The statistical significance of each contribution is evaluated from the change in the negative log-likelihood value when each is included singly, accounting for the change in degrees of freedom. Contributions with significance larger than $5\sigma$ are retained: $K_4^*(2045)^-$, $K^*(1680)^0$, and $K_3^*(1780)^0$ for $\sqrt{s}=3080.00~\text{MeV}$ and $\rho(1450)^+$, $K^*(1680)^-$, $K^*(1680)^0$, $K_2^*(1980)^-$, and $K_2^*(1980)^0$ along with ${\rm NR}_{1^-}$ for $\sqrt{s}=3096.20~\text{MeV}$. The detailed significances and fit fractions (FFs) are listed in Ref.~\cite{prdversion}. One-dimensional projections of the PWA fits on mass spectra are shown in Fig.~\ref{fig:pwaresult}.

The intermediate state configuration obtained from the scan at $\sqrt{s}=3080.00~\text{MeV}$ is adopted for datasets collected at $\sqrt{s}=3000.00$ to $3088.85~\text{MeV}$, while the configuration determined at $\sqrt{s}=3096.20~\text{MeV}$ is used for the remaining energy points. For each energy point, an independent PWA fit is performed with the corresponding fixed intermediate state set, and the complex couplings of all amplitudes are refitted. 

\begin{figure*}[htbp]\centering
  \includegraphics[width=1.0\textwidth]{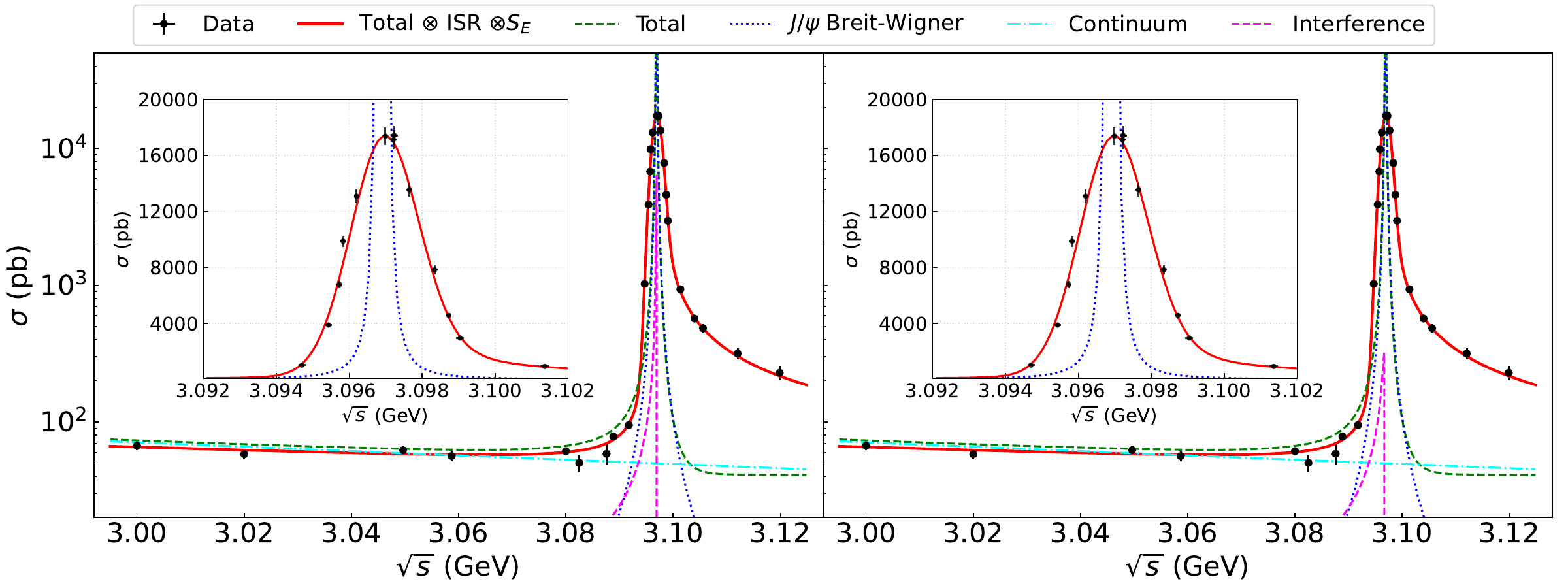}
  \caption{Fits to the observed cross sections of $e^+e^-\to K_S^0 K^+\pi^-$ with positive (left) and negative (right) phase hypotheses.}
  \label{fig:xsfitkskspi}
\end{figure*}

The observed cross sections for $e^+e^-\to K_S^0K^+\pi^-$ and subprocesses are written as
\begin{equation}\label{cs}
  \sigma_{\rm obs}(\sqrt{s})=\frac{N_{\rm{sig}}}{\mathcal{L}\epsilon\mathcal{B}},
\end{equation}
where $N_\text{sig}$ is the signal yield, obtained by counting in the signal region for $e^+e^-\to K_S^0K^+\pi^-$, and for a given subprocess, obtained as $N_\text{sig} = N^{K_S^0K^+\pi^-}_\text{sig} \times \text{FF}_\text{channel}$, where $\text{FF}_\text{channel}$ is the specific channel fit fraction, evaluated using reconstructed-level PHSP MC events, according to the PWA results. $\mathcal{L}$ is the integrated luminosity. The efficiency $\epsilon$ is obtained by weighting the MC simulation according to the PWA results. To account for the ISR and beam energy spread effects, the MC generation with ConExc is iteratively updated using the cross sections obtained from the line shape fit for each channel, until the change in the efficiency between iterations is less than 0.5\%. The symbol $\mathcal{B}$ denotes the product BFs of intermediate states as quoted from the PDG~\cite{pdg}. Corrections for residual data-MC efficiency differences are applied, and further details are provided in Ref.~\cite{prdversion}.

The systematic uncertainties for the cross-section measurements arise from tracking, particle identification, $K_S^0$ reconstruction, luminosity measurement, kinematic fit, radiative correction, input BF, and PWA-related uncertainties. The uncertainties related to the PWA are from the MC model, background description, resonance mass or width, extra resonances, radius parameter, and kinematic model. Detailed descriptions are provided in Ref.~\cite{prdversion}.

A $\chi^2$ fit~\cite{chifit} accounting for correlated and uncorrelated uncertainties across energy points is performed. The fit PDF is described in Eq.~\eqref{eq:fitformular}. Comparing with the partial width of $J/\psi\to \mu^+\mu^-$, the experimental partial decay width of $J/\psi\to f$ is $\Gamma_f=\mathcal{P}(M^2)(\frac{\mathcal{F}}{M^n})^2 \Gamma_{\mu\mu}\left|1+\mathcal{C} e^{i \phi_{\gamma, 3g}}\right|^2$. Changing the partial decay width to BF, we have $\mathcal{B}(J/\psi \to f)=\frac{\Gamma_f}{\Gamma_{\mu\mu}} \mathcal{B}(J/\psi\to \mu\mu)$ and $\Gamma_{ee/\mu\mu}=\Gamma \mathcal{B}(J/\psi\to ee/\mu\mu)$. The free parameters in the line shape fit include the relative magnitude $\mathcal{C}$ and relative phase $\phi_{\gamma, 3g}$ between the strong and EM amplitudes, the BF of the specific decay process $\mathcal{B}(J/\psi\to f)$, and the beam energy spread $S_E$. The resonance mass $M$ and width $\Gamma$ and the leptonic BFs $\mathcal{B}(J/\psi\to ee)$ and $\mathcal{B}(J/\psi\to \mu\mu)$ are fixed to the PDG values~\cite{pdg}. Depending on the ``counting rule''~\cite{heliSele}, the power-law exponent of the form factor $n$ is set to be 3 for $\bar K^0 K^*(892)^0$ and $K^+ K^*(892)^-$ and 5 for $K_S^0 K^+ \pi^-$~\cite{Belle:2013hkg,BaBar:2007ceh}. The relative phase between the continuum and EM amplitudes $\phi_{\gamma, \text{cont}}$ was predicted to be zero by pQCD and confirmed by Ref.~\cite{wangyadi}. Therefore, we fix it to zero across all channels. Minimization is performed using the TMinuit package~\cite{James:1975dr}, and uncertainties are estimated using the HESSE algorithm.

\begin{table*}[htbp]\centering
  \caption{Best-fit results to the observed cross sections $e^+e^-\to K_S^0 K^+\pi^-$ and its subprocesses. The uncertainties include both statistical and systematic effects.}
  \begin{tabular}{c|c|c|c|c|c|c}
    \toprule
    Channel & $\phi_{\gamma, 3g}$ sign & $\mathcal{C}$ & $\mathcal{B}~(\times 10^{-3})$ & $\phi_{\gamma, 3g}~(^\circ)$ & $S_E$ (MeV) & $\chi^2/\text{n.d.f.}$ \\
    \midrule
    \multirow{2}{*}{$K_S^0K^+\pi^-$} &
    Positive   & $4.31\pm 0.22$  & $5.17\pm 0.20$                 & $ 123.7\pm 5.3$  & $0.90\pm 0.02$ & $16.0/22$           \\
    & Negative & $4.38\pm 0.22$  & $5.36\pm 0.20$                 & $-123.1\pm 5.2$  & $0.90\pm 0.02$ & $16.0/22$           \\
    \midrule
    \multirow{2}{*}{$\bar{K}^0K^*(892)^0$} &
    Positive   & $3.80\pm 0.26$  & $4.62\pm 0.19$                 & $ 147.0\pm 11.7$ & $0.92\pm 0.03$ & $32.6/22$           \\
    & Negative & $3.85\pm 0.25$  & $4.80\pm 0.19$                 & $-146.0\pm 11.6$ & $0.92\pm 0.03$ & $32.6/22$           \\
    \midrule
    $ K^+K^*(892)^-$ & ...  & $28.66\pm 2.68$ & $6.41 \pm 0.25$     & $180.2\pm 35.2$  & $0.89\pm 0.02$ & $18.9/22$  \\
    \bottomrule
    \end{tabular}
	\label{tab:xsparakskspi}
\end{table*}

There are two sets of parameters that describe the cross section equally well. Those with the parameters $\phi_{\gamma,3g}\in(0,180)^\circ$ and $\phi_{\gamma,3g}\in(-180,0)^\circ$ are called the positive and negative solutions, respectively. Current data cannot distinguish between the positive and negative solutions because the squared amplitude has an intrinsic twofold ambiguity~\cite{Bai:2019jrb}. The best-fit results of $J/\psi\to K_S^0 K^+\pi^-$ and its subprocesses $J/\psi\to \bar{K}^0 K^*(892)^0$ and $J/\psi\to K^+ K^*(892)^-$ are shown in Figs.~\ref{fig:xsfitkskspi}, \ref{fig:xsfitk0kstar}, and \ref{fig:xsfitkkstar}, and the fitted parameters are summarized in Table~\ref{tab:xsparakskspi}. For the neutral channel, $|\phi_{\gamma,3g}|\simeq(147\pm 12)^\circ$ implies a sizable dispersive strong component opposite in sign to the EM amplitude. The $4.8\sigma$ deviation from $90^\circ$ shows that strong-EM orthogonality is not universal~\cite{ppphase,NNphase,wangyadi,sunzequn}, disfavoring scenarios with a channel-independent, predominantly imaginary strong amplitude, such as a universal vector glueball admixture or absorptive OZI contributions~\cite{gluonball1,gluonball2,OZIamp}, while remaining compatible with FSI dynamics~\cite{FSI1,FSI2,FSIopechar1,FSIopechar2}.

The systematic uncertainties on the parameters arise from two sources. The first is the parameterization of the continuum, which is evaluated by performing alternative fits with $n$ varied by $\pm 2$. The second comes from the fixed parameters of the resonance state: $M$, $\Gamma$, $\mathcal{B}(J/\psi\to ee)$, and $\mathcal{B}(J/\psi\to \mu\mu)$, which are estimated by shifting each value within $\pm 1\sigma$. Detailed descriptions are provided in Ref.~\cite{prdversion}.

\begin{figure}[htbp]\centering
  \includegraphics[width=0.48\textwidth]{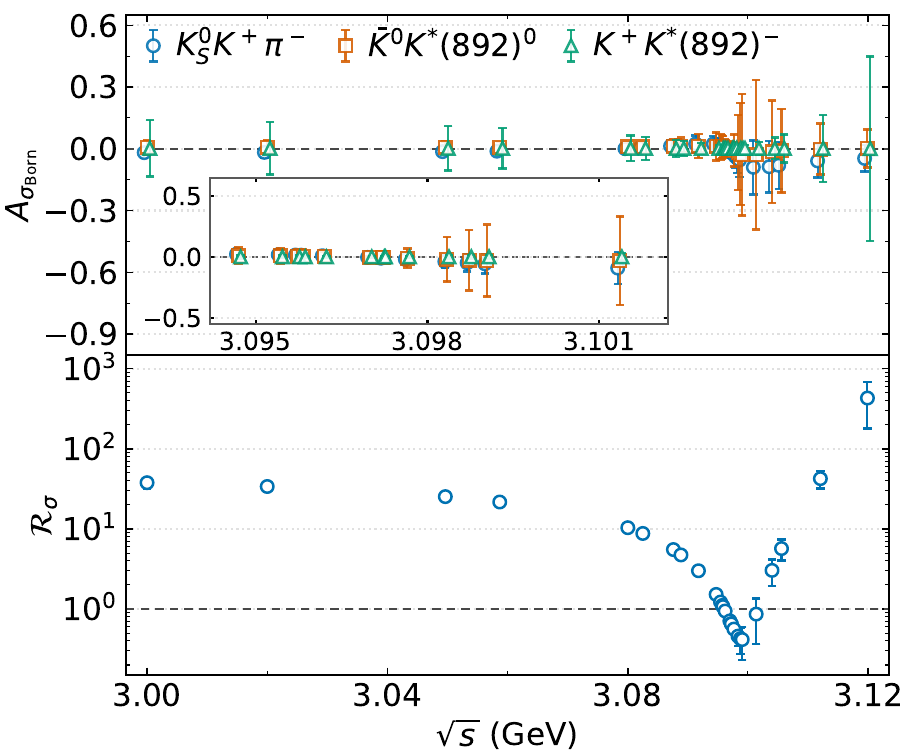}
  \caption{Asymmetry observables of the Born cross sections (top) and the $\mathcal{R}_\sigma$ of the Born cross sections (bottom).}
  \label{fig:asy}
\end{figure}

The isospin-breaking effect is reflected in the measured ratio $\mathcal{R}_{K^*\bar{K}}=0.723\pm 0.013$ or $0.752\pm 0.014$. These ratios are obtained from a simultaneous $\chi^2$ fit to the neutral and charged modes. When only the strong amplitude is considered, the ratio increases to $\mathcal{R}^{3g}_{K^*\bar{K}}=1.062\pm 0.070$ or $1.090\pm 0.063$, consistent with unity within $1.4\sigma$. By contrast, the contribution from the EM amplitude alone yields a much larger ratio of $\mathcal{R}^{\gamma}_{K^*\bar{K}}=62.0\pm 12.2$, indicating a very pronounced isospin-breaking effect from the EM interaction. In the energy region dominated by continuum production, far from the $J/\psi$ resonance, the ratio approaches $\mathcal{R}_\sigma=37.74\pm 5.57$, as shown in the bottom panel of Fig.~\ref{fig:asy}. The $\mathcal{C}=3.80\pm 0.26$ versus $28.66\pm 2.68$ corresponds to $|\mathcal{A}_{\gamma}/\mathcal{A}_{3g}|=0.263\pm 0.018$ and $0.035\pm 0.003$ in the neutral and charged modes, respectively. Together with $\mathcal{R}^{3g}_{K^*\bar K}$ and $\mathcal{R}^{\gamma}_{K^*\bar K}$, these results indicate that the isospin breaking resides mainly in the EM amplitude, while the strong contribution remains isospin symmetric. This qualitatively supports effective Lagrangian expectations of strong dominance and similar strong flavor breaking effects in both modes~\cite{VPEFT}, while the pronounced observed EM amplitude constrains the channel-dependent EM operators and possible hadronic loop effects~\cite{FSIopechar1,FSIopechar2}. Direct $CP$ asymmetries are extracted both from the Born cross sections and from the line shape fit parameters, as shown in the top panel of Fig.~\ref{fig:asy} ($A_{\sigma_{\rm Born}}$, for example), while the parameter asymmetries are summarized in Table~\ref{tab:param_asym}. All measured null-test observables are consistent with the standard model predictions within $1\sigma$. The $CP$ results are quoted for the positive phase solution; the negative phase solution gives the same $CP$-null conclusion.

\begin{table}[htbp]\centering
  \caption{Parameter asymmetries for $\mathcal{C}$, $\mathcal{B}$, and $\phi_{\gamma,3g}$ in cross-section fits.}
  \resizebox{\linewidth}{!}{
    \begin{tabular}{cccc}
      \toprule
      Channel                         & $A_{\mathcal C}$ & $A_{\mathcal B}$& $\Delta\phi_{\gamma,3g}~(^\circ)$  \\
      \midrule
      $K_S^0K^+\pi^-$       & $0.018\pm 0.027$  & $0.001\pm 0.003$  & $4.7\pm 7.3$   \\
      $\bar{K}^0K^*(892)^0$ & $0.001\pm 0.048$  & $0.002\pm 0.009$  & $1.7\pm 15.8$  \\
      $K^+K^*(892)^-$       & $-0.000\pm 0.087$ & $0.002\pm 0.008$  & $-0.0\pm 66.7$ \\
      \bottomrule
    \end{tabular}}
  \label{tab:param_asym}
\end{table}

In summary, based on a direct line shape scan with unprecedented energy granularity, we report the first extraction of the relative magnitude $\mathcal{C}$ and phase $\phi_{\gamma,3g}$ between the strong and EM amplitudes in $J/\psi\to K^*(892)\bar K$. The corresponding BFs, determined from coherent line shape fits, are consistent with the world averages with substantially improved precision. The strong-EM phase is found to be dependent with decay modes, and the $J/\psi\to\bar K^0K^*(892)^0$ channel disfavors the orthogonal relation by $4.8\sigma$, a relative real amplitude ($0^\circ$) by $12.6\sigma$, and $180^\circ$ by $2.8\sigma$, while that of $J/\psi\to K^+K^*(892)^-$ is consistent with $180^\circ$ within uncertainties. The departure from $90^\circ$ demonstrates that strong-EM orthogonality is not universal and implies a sizable dispersive strong component opposite to the EM amplitude. This challenges simple descriptions based on a channel-independent, predominantly imaginary strong amplitude and provides a quantitative target for FSI dynamics. The large $\mathcal{C}$ value in the charged mode implies a highly suppressed EM contribution; together with the observed phase pattern, this marks channel dependence arising predominantly from the EM amplitude. Separating strong and EM components further shows that the strong amplitude is approximately isospin symmetric. The first search for direct $CP$ violation in $J/\psi\to K^*(892)\bar K$ decays finds no evidence in either the measured Born cross sections or the fitted amplitude parameters. These results provide new constraints on the phase structure and isospin systematics in $J/\psi$ decays and will help discriminate among candidate long-distance mechanisms with improved continuum measurements and amplitude analyses.

\textbf{Acknowledgement}

The BESIII Collaboration thanks the staff of BEPCII (https://cstr.cn/31109.02.BEPC) and the IHEP computing center for their strong support. This work is supported in part by National Key R\&D Program of China under Contracts Nos. 2023YFA1606000, 2023YFA1606704, 2025YFA1613900; National Natural Science Foundation of China (NSFC) under Contracts Nos. 11635010, 11935015, 11935016, 11935018, 12025502, 12035009, 12035013, 12061131003, 12192260, 12192261, 12192262, 12192263, 12192264, 12192265, 12221005, 12225509, 12235017, 12342502, 12361141819, 12535005; the Chinese Academy of Sciences (CAS) Large-Scale Scientific Facility Program; the Strategic Priority Research Program of Chinese Academy of Sciences under Contract No. XDA0480600; CAS under Contract No. YSBR-101; 100 Talents Program of CAS; The Institute of Nuclear and Particle Physics (INPAC) and Shanghai Key Laboratory for Particle Physics and Cosmology; ERC under Contract No. 758462; German Research Foundation DFG under Contract No. FOR5327; Istituto Nazionale di Fisica Nucleare, Italy; Knut and Alice Wallenberg Foundation under Contracts Nos. 2021.0174, 2021.0299, 2023.0315; Ministry of Development of Turkey under Contract No. DPT2006K-120470; National Research Foundation of Korea under Contract No. NRF-2022R1A2C1092335; National Science and Technology fund of Mongolia; Polish National Science Centre under Contract No. 2024/53/B/ST2/00975; STFC (United Kingdom); Swedish Research Council under Contract No. 2019.04595; U. S. Department of Energy under Contract No. DE-FG02-05ER41374

\textbf{DATA AVAILABILITY}

The data that support the findings of this article are not publicly available upon publication because it is not technically feasible and/or the cost of preparing, depositing, and hosting the data would be prohibitive within the terms of this research project. The data are available from the authors upon reasonable request.

\onecolumngrid
\appendix
\section{End Matter}
\section{Figures of fit results of the cross sections of $e^+e^-\to\bar{K}^0K^*(892)^0$ and $e^+e^-\to K^+K^*(892)^-$}

\begin{figure*}[htbp]\centering
  \includegraphics[width=1.0\textwidth]{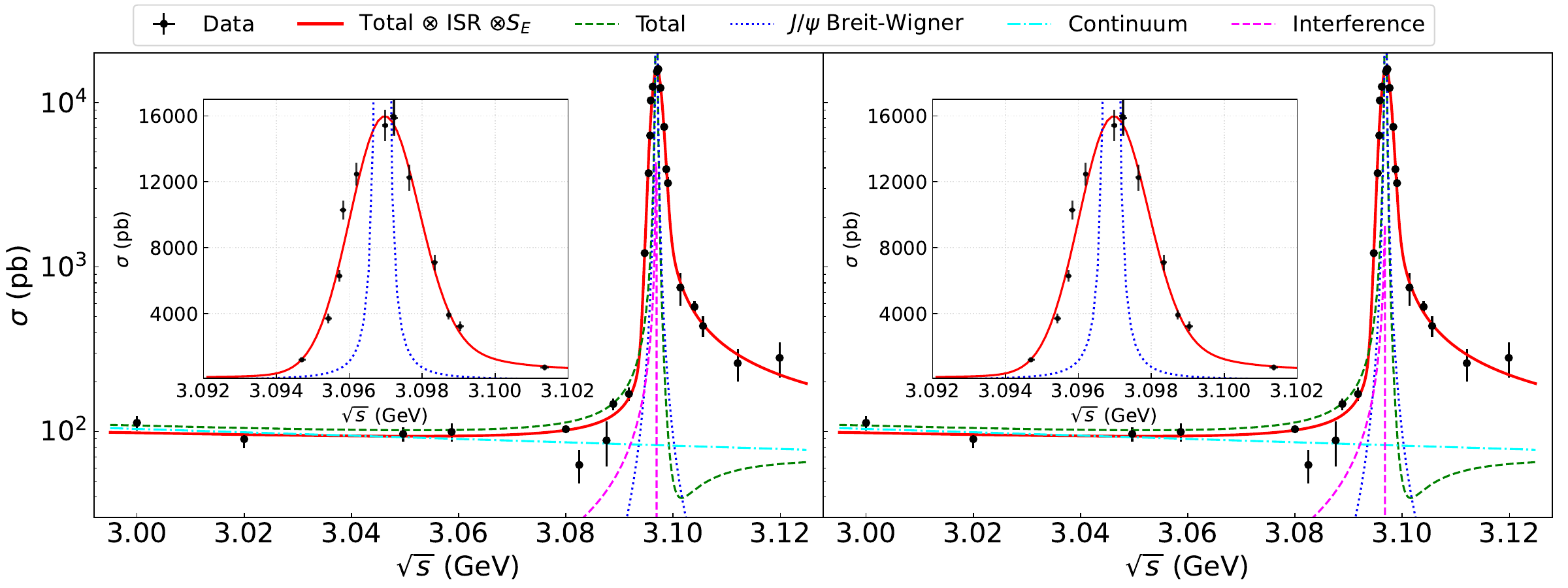}
  \caption{Fits to the observed cross sections of $e^+e^-\to \bar{K}^0K^*(892)^0$ with positive (left) and negative (right) phase hypotheses.}
  \label{fig:xsfitk0kstar}
\end{figure*}

\begin{figure*}[htbp]\centering
  \includegraphics[width=0.6\textwidth]{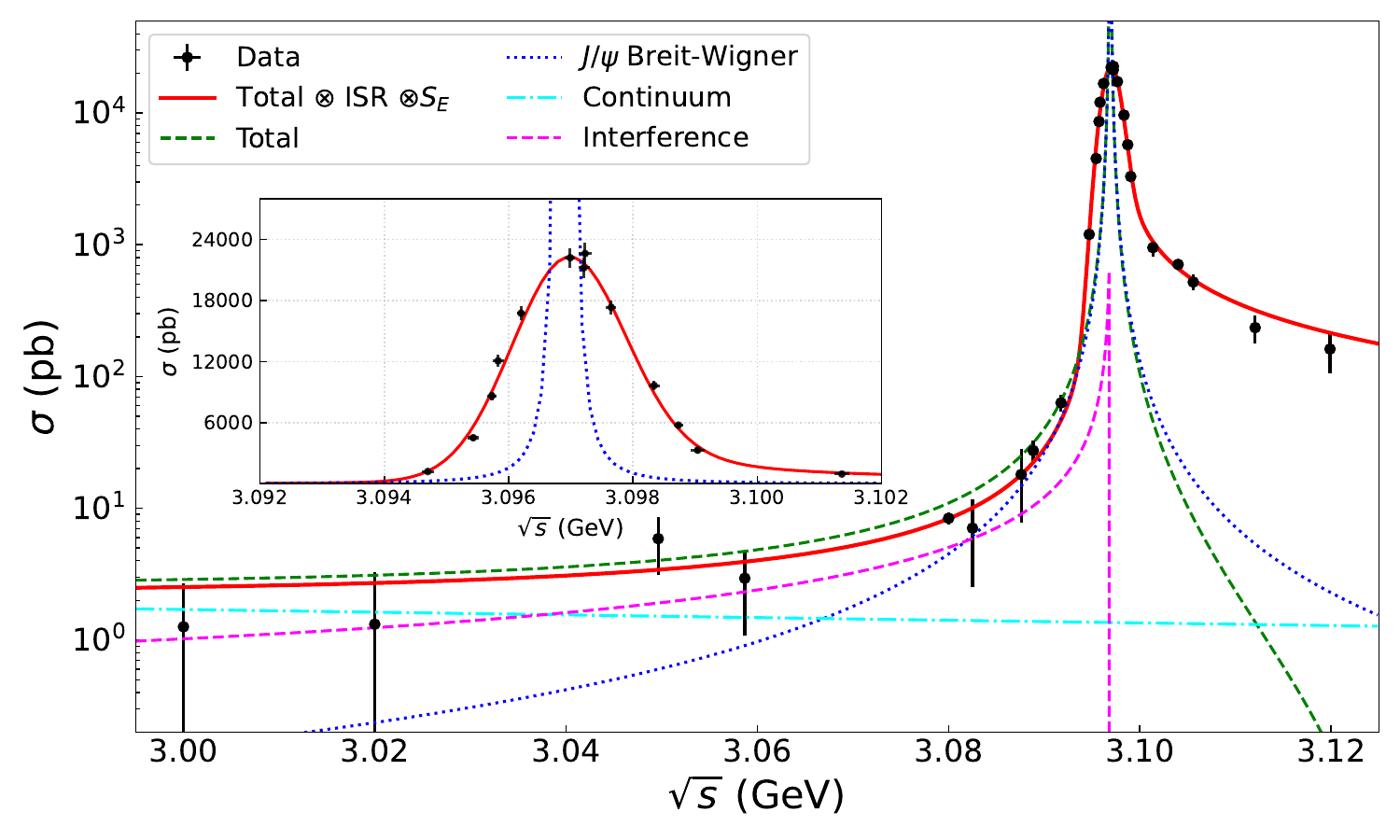}
  \caption{Fits to the observed cross sections of $e^+e^-\to K^+K^*(892)^-$.}
  \label{fig:xsfitkkstar}
\end{figure*}

\end{document}